\documentclass[pra,aps,superscriptaddress,floatfix,tightenlines,twocolumn]{revtex4-1}

\usepackage{graphicx}
\usepackage{dcolumn}   % needed for some tables
\usepackage{bm}        % for math
\usepackage{amssymb}   % for math
\usepackage{amsmath}
\usepackage{url,color}
\usepackage{layout}
\usepackage{epsfig}
\usepackage{epstopdf}
\usepackage{booktabs}
\usepackage{float,CJK}
\usepackage{array}
\usepackage{textpos}

\usepackage[hyperindex,breaklinks]{hyperref}
\begin{document}

%\preprint{OU-HET ????}

\title{Neural ODE and Holographic QCD}

\author{Koji Hashimoto}%
\email{koji@phys.osaka-u.ac.jp}
\affiliation{%
 Department of Physics, Osaka University, 
  Toyonaka, Osaka 560-0043, Japan
}%
\author{Hong-Ye Hu}%%
\email{hyhu@ucsd.edu }
\author{Yi-Zhuang You}%%
\email{yzyou@physics.ucsd.edu}
\affiliation{%
Department of Physics, University of California at San Diego, La Jolla, CA 92093, USA}%

\begin{CJK}{UTF8}{gbsn}

\begin{abstract}
The neural ordinary differential equation (Neural ODE) is a novel machine learning architecture whose weights are smooth functions of the continuous depth. We apply the Neural ODE to holographic QCD by regarding the weight functions as a bulk metric, and train the machine with lattice QCD data of chiral condensate at finite temperature. The machine finds consistent bulk geometry at various values of temperature and discovers the emergent black hole horizon in the holographic bulk automatically. The holographic Wilson loops calculated with the emergent machine-learned bulk spacetime have consistent temperature dependence of confinement and Debye-screening behavior. In machine learning models with physically interpretable weights, the Neural ODE frees us from discretization artifact leading to difficult ingenuity of hyperparameters, and improves numerical accuracy to make the model more trustworthy. 

\end{abstract}

%\pacs{}

\maketitle
\end{CJK}

%%%%%%%%%%%%%%%%%%%%%%%%%%%%%%%%%%%%%%%%%%%%%%%%%%%%%%%%%%%%%%%%%%%%%%%%%%%%%%%%%%%%%%%%%
\section{Introduction}

Applying machine learning to solve physics problems\cite{Carleo:2019ptp, Ruehle:2020jrk} has generated a growing research interest in recent years. Machine learning holography is an emerging direction in this field, which introduces artificial intelligence to discover the holographic bulk theory behind generic quantum systems on the holographic boundary. Multiple approaches have been developed to capture different aspects of the holographic duality\cite{Gan:2017xy, Hashimoto:2018bnb, Hashimoto:2018ftp, You:2017guh,2019arXiv190300804H, Hashimoto:2019bih, Han2020Deep, Akutagawa:2020yeo}. For example, the entanglement feature learning (EFL)\cite{You:2017guh} can establish the emergent holographic spacial geometry simply from the entanglement entropy data on the holographic boundary. The anti-de Sitter / deep learning (AdS/DL) correspondence takes a different approach\cite{Hashimoto:2018bnb,Hashimoto:2018ftp,Hashimoto:2019bih,Akutagawa:2020yeo} by implementing the holographic principle\cite{Maldacena:1997re,Gubser:1998bc,Witten:1998qj} in a deep neural network, where the neural network is regarded as the classical equation of motion for propagating fields on a discretized curved spacetime. Further progress has been made by the neural network renormalization group (Neural RG)\cite{2019arXiv190300804H}, which learns to construct the exact holographic mapping between the boundary and the bulk field theories at the partition function level. All these approaches share a common theme that the emergent dimension of the holographic bulk corresponds to the depth dimension of the deep neural network, and the neural network itself is regarded as the bulk spacetime. As the neural network learns to interpret the holographic boundary data serving from its input layer, the network weights in deeper layers get optimized, which then leads to the optimal holographic bulk description for the boundary data.

However, the development so far has been based on the discretization of the holographic bulk dimension, because the neural network layers are intrinsically discrete in typical deep learning architectures. It is desired to make this dimension continuous, as a smooth holographic spacetime is physically required in the classical limit. In this work, we explore this possibility, based on the recent development of the neural ordinary differential equation (Neural ODE)\cite{chen2018neural} approach. The Neural ODE is a generalization of the deep residual network\cite{He2015Deep} to a continuous-depth network with the network weights replaced by a continuous function. It provides a trainable model of differential equations that can evolve the initial input to the final output continuously. The Neural ODE is particularly suitable for the AdS/DL approach because the goal here is precisely to infer the differential equation that describes the propagation of the bulk field in a continuous space-time with smooth geometry. In this context, the continuous network weights of the Neural ODE have a physical interpretation related to the metric function that characterizes the curved spacetime in the holographic bulk. An interpretable spacetime geometry emerges as the neural network is trained, which demonstrate a scenario of machine-assisted discovery in theoretical physics, where the artificial intelligence plays a more active role in the scientific process other than a tool for data processing.

The AdS/DL applied to holographic QCD would be a nice ground to test the effectiveness of the Neural ODE in physics applications. The Neural ODE brings to us two advances: the removal of artificial regularizations and the improvement of accuracy. In previous works \cite{Hashimoto:2018ftp, Hashimoto:2018bnb, Yan:2020wcd, Akutagawa:2020yeo}, due to the discrete nature of the neural network, technical regularization terms are introduced to remove the discretization artifacts and to ensure the smoothness of the network weights.\footnote{See Ref.~\cite{Hashimoto:2019bih} for the physical meaning of the regularization as an Einstein action.} Such regularization is no longer needed in the Neural ODE approach. Furthermore, for the network to be identified with a field equation in the curved spacetime, the Euler method for the ordinary differential equation was introduced for simplicity, though the Euler integration generically suffers from large numerical errors. Replacing the discrete neural network with the Neural ODE provides a natural interpretation of the metric function in the smooth spacetime, and at the same time, would greatly enhance the accuracy. The improved accuracy of the Neural ODE is simply due to the advanced ODE solver equipped in the Neural ODE framework. The discretization along the integrated coordinate is optimized adaptively, rather than given ad hoc as hyperparameters. This is especially useful when the metric function contains coordinate singularity at the black hole horizon. The required accuracy depends on the purpose and the method of how machine learning is applied.\footnote{For example, a hybrid version \cite{nagai2019self} of the self-learning Monte Carlo \cite{liu2017self} is a novel way to train effective Hamiltonian while keeping the desired accuracy.} In our present case of the AdS/DL, as is explicitly shown, the accuracy improvement is sufficient for exploring emergent geometries at various values of temperature.

In this paper, following the holographic QCD framework of Ref.~\cite{Hashimoto:2018bnb}, we use the Neural ODE to find bulk spacetimes emergent out of the given data of chiral condensate of lattice QCD. The Neural ODE not only discovers a spacetime which is consistent with that of Ref.~\cite{Hashimoto:2018bnb}, but also greatly enhances the power of machine learning method. The emergent geometry turns out to incorporate automatically the presence of the black hole horizon, and the Neural ODE enables us to further explore geometries for different values of temperature, with improved accuracy. The temperature dependence of holographic Wilson loops, calculated by the emergent geometry trained with the Neural ODE, turns out to coincide qualitatively with the known lattice QCD results of the Wilson loops. Interestingly, we find that the radial derivative of the volume factor of the emergent geometry does not depend on the temperature, and the temperature dependence of the chiral condensate solely stems from that of the bulk scalar coupling constant.

The organization of this paper is as follows.
In Sec.~\ref{sec:2}, we briefly review the holographic QCD framework adopted in Ref.~\cite{Hashimoto:2018bnb} and the Neural ODE \cite{chen2018neural}. 
In Sec.~\ref{sec:3}, we apply the Neural ODE to train the machine (which is equivalent to the holographic QCD system) and find emergent geometry for various values of the temperature. 
In Sec.~\ref{sec:4}, we introduce a way to calculate consistent full components of the metric from
the emergent volume factor, with which we calculate holographic Wilson loops. They qualitatively agree with Wilson loops evaluated in lattice QCD. 
Sec.~\ref{sec:5} is for a summary and discussions. Appendix \ref{sec:A} is about details of the Neural ODE. %, and appendix \ref{sec:B} is multi-temperature data training results.

%%%%%%%%%%%%%%%%%%%%%%%%%%%%%%%%%%%%%%%%%%%%%%%%%%%%%%%%%%%%%%%%%%%%%%%%%%%%%%%%%%%%%%%%%
\section{Review: AdS/CFT model and Neural ODE}
\label{sec:2}

%%%%%%%%%%%%%%%%%%%%%%%%%%%%%%%%%%%%%%%%%%%%%%%%%%%%%%%%%%%%%%%%%%%%%%%%%%%%%%%%%%%%%%%%%
\subsection{Bulk field theory}

The holographic principle \cite{Maldacena:1997re,Gubser:1998bc,Witten:1998qj}, also known as AdS/CFT correspondence, is a profound relation between 
a $d$-dimensional quantum field theory (QFT) and a $(d+1)$-dimensional gravity theory. It has been successfully applied to a large class of strongly coupled QFTs in high energy theory and condensed matter theory. Despite its success, a constructive way of finding the holographic gravity dual theory for a given QFT is lacking. If we have the experimental response data of a quantum system under external probing fields, can we model it holographically by a classical field theory in some curved
geometry? The entanglement feature learning \cite{You:2017guh, Vasseur:2018gfy} and the 
AdS/DL correspondence \cite{Hashimoto:2018ftp,Hashimoto:2018bnb} can answer that question
in a concrete setup. Here we briefly review the setup of Ref.~\cite{Hashimoto:2018bnb}, for which we apply
the Neural ODE method in later sections.

We assume the $d+1$-dimensional bulk spacetime coordinated by $(t,\eta, x_1, \cdots, x_{d-1})$ including the time dimension $t$, the space dimensions $x_i$ and the holographic bulk dimension $\eta$. We assume the translation symmetry except for the $\eta$ direction, and the spacial homogeneity in $(x_1,\cdots,x_{d-1})$, then in the gauge $g_{\eta\eta}=1$, 
the holographic bulk spacetime can be described by the following metric (we will consider $d=4$ specifically)
\begin{equation}
\mathrm{d}s^{2} = -f(\eta)\mathrm{d}t^{2}+\mathrm{d}\eta^{2}+g(\eta)(\mathrm{d}x^{2}_{1}+\cdots+\mathrm{d}x^{2}_{d-1}) \, .
\label{metricd+1}
\end{equation}
The dual quantum field theory lives in a $d$-dimensional flat spacetime spanned by $(t, x_1, \cdots,x_{d-1})$ on the holographic boundary. 
We call $\eta$ the radial coordinate and the others are angular directions. The spacetime volume factor is
\begin{equation}
\sqrt{|g|}=\sqrt{-\det g}=\sqrt{f(\eta)g(\eta)^{d-1}} \, .
\end{equation}
A scalar field $\phi$ in this curved spacetime is described by the action:
\begin{equation}
\begin{split}
S[\phi] = \dfrac{1}{2}\int \sqrt{|g|}\left( g^{\mu \nu}\partial_{\mu}\phi \partial_{\nu}\phi+m^{2}\phi^2+\frac{\lambda}{2}\phi^4\right) \, .
%\\& =\dfrac{1}{2}\int d^{d+1}x \sqrt{|g|}\left(\dfrac{-1}{\sqrt{g}}\phi\, \partial_{\mu}\sqrt{|g|} g^{\mu\nu}\partial_{\nu}\phi+ m^{2}\phi^{2}+\frac{\lambda}{2}\phi^4\right).
\end{split}
\end{equation}
The saddle point equation (the classical equation of motion) $\delta S/\delta \phi=0$ reads,
\begin{equation}\label{eq: EoM}
-\dfrac{1}{\sqrt{|g|}}\partial_{\mu}\left(\sqrt{|g|}g^{\mu\nu}\partial_{\nu}\phi \right)+m^{2}\phi+\lambda \phi^3=0 \, .
\end{equation}
Since we are interested in homogeneous static condensate in the dual quantum field theory, we assume that $\phi$ is only a function of $\eta$. Then
%Suppose we only consider the static and uniform $g^{\mu\nu}$. We can neglect the space-time derivative of $g^{\mu\nu}$, and 
Eq.~\eqref{eq: EoM} becomes,
\begin{equation}
\begin{aligned}
-\partial_{\eta}^{2}\phi-(\partial_{\eta}\ln \sqrt{|g|})\partial_{\eta}\phi+ m^{2}\phi+\lambda \phi^3=0
\end{aligned}
\end{equation}
or equivalently, we could write it as
\begin{equation}
\begin{aligned}
&\pi = \partial_{\eta}\phi \, ,\\
&\partial_{\eta}\pi+h(\eta)\pi -m^{2}\phi -\lambda \phi^3 = 0 \, ,\label{EOM}
\end{aligned}
\end{equation}
where the metric function is (with $d=4$)
\begin{equation}
h(\eta) \equiv \partial_{\eta}\ln\sqrt{f(\eta)g(\eta)^{d-1}} \, .
\label{hdef}
\end{equation}
The input data is the pair $\phi(\eta \sim \infty), \pi(\eta\sim \infty)$ near the AdS horizon. And the field will propagate following the classical equation of motion Eq.~\eqref{EOM}. On the other hand, there is black hole horizon at $\eta\sim 0$. The on-shell static scalar field satisfies the black hole boundary condition
\begin{equation}
\left[\dfrac{2}{\eta}\pi -m^{2}\phi -\lambda \phi^3\right]_{\eta\sim 0}=0 \, ,
\end{equation}
or equivalently, we could require
\begin{align}
  \pi(\eta\sim 0)=0 \, .
\label{blackhole_condition}
\end{align}

\begin{figure*}[t]
\centering
\includegraphics[width = 0.4\linewidth]{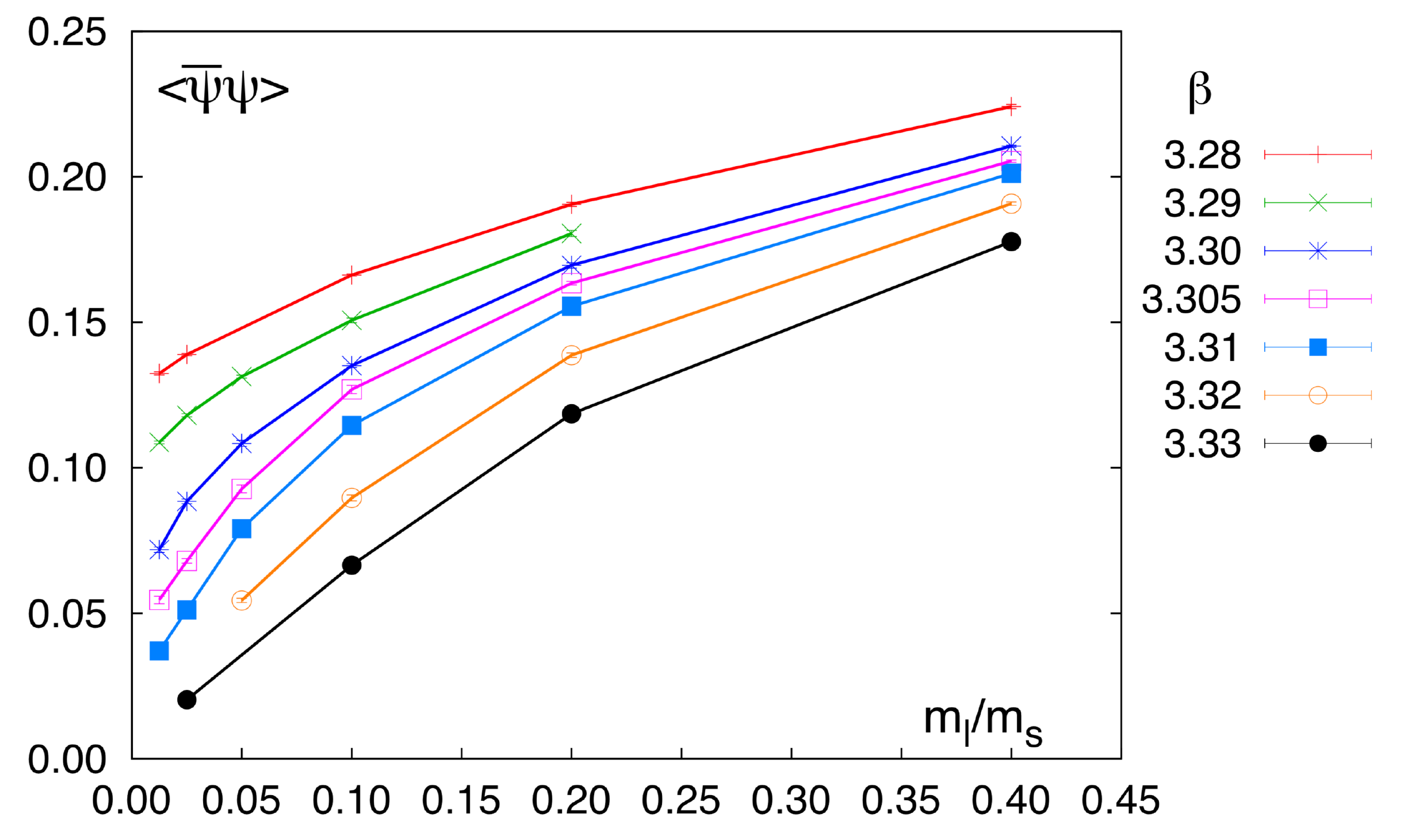}
\includegraphics[width = 0.45\linewidth]{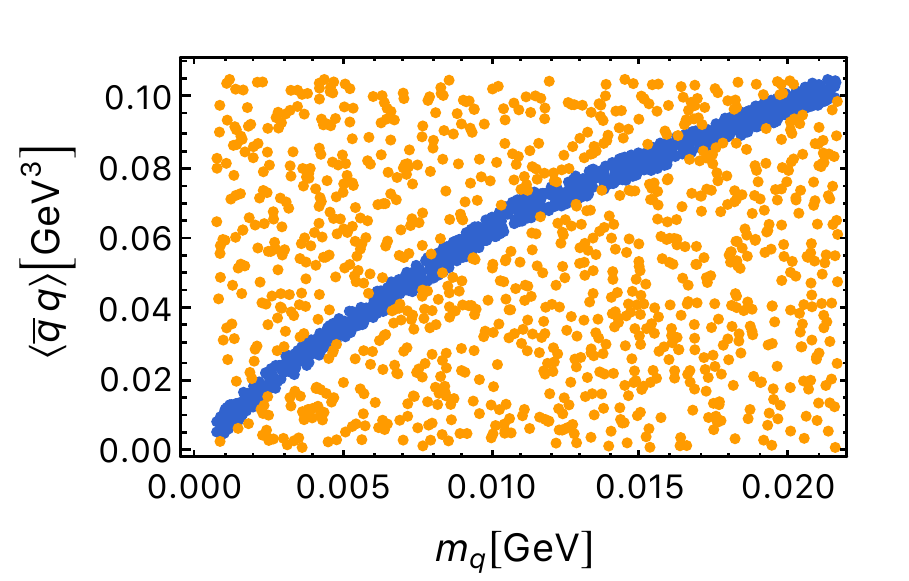}
\caption{
Left: 
The lattice QCD data plot for the chiral condensate as a function of quark mass, for various values of the temperature (excerpt from \cite{Karsch:2008ch}. The horizontal axis is the light quark mass normalized by the strange quark mass.
Right:
Data used for training. The orange dots are negative data, and the blue dots are positive data.
}
\label{fig:data}
\end{figure*}

The mapping between the asymptotic value of the scalar field $\phi(\eta \sim\infty)$ and the data of the dual quantum field theory is given by the AdS/CFT dictionary with the asymptotically AdS spacetime with the AdS radius $L$ \cite{Hashimoto:2018bnb},
\begin{equation}\label{eq: asymptotic}
  L^{3/2}\phi\sim \alpha e^{-\eta/L}
  + \beta e^{-3\eta/L}-\frac{\lambda \alpha^3}{2L^2}\eta e^{-3\eta/L}
\end{equation}
for an operator ${\cal O}$ whose dimension is three, corresponding to the bulk scalar field $\phi$ with the mass $m^2 = -3/L^2$. The coefficients are related to the condensate as
\begin{align}
  \alpha = \frac{\sqrt{N_c}}{2\pi}m_{\cal O} \, , \quad
  \beta = \frac{\pi}{\sqrt{N_c}}\langle {\cal O}\rangle L^3 \, .
\end{align}
Here, $m_{\cal O}$ is the source for the operator ${\cal O}$ of the quantum field theory, and $N_c$ denotes the color number in QFT and hence we set $N_c=3$ as we focus on QCD later.
Therefore, the data of one-point function of the quantum field theory $\{m_{\cal O},{\langle\cal O\rangle}\}$ is given, it is mapped to on-shell configuration of $\phi(\eta)$ and $\pi(\eta)$ (by taking derivative on both sides of Eq.~\eqref{eq: asymptotic}) near the holographic boundary $\eta\sim\infty$.

The experimental data pairs ($\phi(\eta\sim \infty), \pi(\eta\sim \infty)$) can be viewed as the positive data. And they will satisfy the black hole boundary condition Eq.~\eqref{blackhole_condition}) after following the classical equation of motion Eq.~\eqref{EOM}. We could also view pairs of data $\phi(\eta\sim \infty), \pi(\eta\sim \infty)$ that does not lie on the experimental curve as negative data. We expect those negative data will not satisfy the black hole boundary condition. Therefore, this becomes a binary classification problem, with the propagation equation Eq.~\eqref{EOM}. Here, for a given data of the condensate, the parameters in the differential equation to be learned are:
the continuous metric function $h(\eta)$, the AdS radius $L$ and interaction coupling $\lambda$ are in general unknown. 

We regard Eq.~\eqref{EOM} as a neural network,
and the network weights are the metric function and other parameters. For that purpose, the numerical method known as the Neural ODE is a perfect framework to find the optimal estimation for those unknown parameters. In the following, we will briefly review the Neural ODE method.

%%%%%%%%%%%%%%%%%%%%%%%%%%%%%%%%%%%%%%%%%%%%%%%%%%%%%%%%%%%%%%%%%%%%%%%%%%%%%%%%%%%%%%%%%
\subsection{Neural ODE}

The Neural ODE \cite{chen2018neural} is a novel framework of deep learning. Instead of mapping the input to the output by a set of discrete layers, the Neural ODE evolves the input to the output by a differential equation, which is trainable. The general form of the differential equation reads
\begin{equation}
\dfrac{\mathrm{d}z(t)}{\mathrm{d}t} = f_{\theta}(z(t), t) \, ,
\label{NODE}
\end{equation}
where the vector $z$ denotes the collection of hidden variables and $\theta$ denotes all the trainable parameters (which could also be $t$-dependent) in the neural network. Without loss of generality, suppose we have observations at the beginning and end of the trajectory: $\{(z_{0}, t_{0}), (z_{1}, t_{1})\}$. One starts the evolution of the system from $(z_{0}, t_{0})$ for time $t_{1}-t_{0}$ with parameterized velocity function $f_{\theta}(z(t), t)$ using any ODE solver. Then the system will end up at a new state $(z_{1}, t_{1})$. Formally, we could consider optimizing the general loss function $\mathcal{L}$, which explicitly depends on the output $z_1$ as 
\begin{equation}
\mathcal{L}(z_{1})=\mathcal{L}\left(\int_{t_{0}}^{t_{1}}\mathrm{d}t~f_{\theta}(z(t),t)\right) \, .
\end{equation}
To back-propagate the gradient with respect to the parameters $\theta$, one introduces the adjoint parameters $a(t) = \frac{\partial \mathcal{L}}{\partial {z}(t)}$ and their corresponding backward dynamics,
\begin{equation}\label{NODE back}
\dfrac{\mathrm{d}{a}(t)}{\mathrm{d}t} = -{a}(t)\cdot\dfrac{\partial {f}_\theta}{\partial {z}} \, .
\end{equation}
After solving Eqs.~\eqref{NODE} and \eqref{NODE back} jointly, the parameter gradient can be evaluated from
%We could find the gradients by solving the following adjoint ODE equations:
%\begin{equation}
%\dfrac{\partial \mathcal{L}}{\partial \vec{z}(t_{0})} = \int^{t_{0}}_{t_{1}}\vec{a}_{z}(t)\dfrac{\partial \vec{f}}{\partial \vec{z}}dt
%\label{eq1}
%\end{equation}
\begin{equation}
\dfrac{\partial \mathcal{L}}{\partial \theta} = \int^{t_{0}}_{t_{1}}a(t)\cdot\dfrac{\partial f_\theta}{\partial \theta}\mathrm{d}t \, .
\label{NODE grad}
\end{equation}
The derivation Eq.~\eqref{NODE grad} can be found in the appendix.

%%%%%%%%%%%%%%%%%%%%%%%%%%%%%%%%%%%%%%%%%%%%%%%%%%%%%%%%%%%%%%%%%%%%%%%%%%%%%%%%%%%%%%%%%
\section{Emergent spacetime from Neural ODE}
\label{sec:3}

%%%%%%%%%%%%%%%%%%%%%%%%%%%%%%%%%%%%%%%%%%%%%%%%%%%%%%%%%%%%%%%%%%%%%%%%%%%%%%%%%%%%%%%%%
\subsection{Learning architecture}

%%%%%%%%%%%%%%%%%%%%%%%%%%%%%%%%%%%%%%%%%%%%%%%%%%%%%%%%%%%%%%%%%%%%%%%%%%%%%%%%%%%%%%%%%
\subsubsection{Neural ODE and bulk equation}

In the form of the first order differential equation, the equations of motion for the bulk field Eq.~\eqref{EOM} can be translated to the Neural ODE Eq.~\eqref{NODE} by the following identifications:
\begin{align}
  (\pi, \phi) \leftrightarrow {z} \, ,
  \quad \eta \leftrightarrow t \, .
\end{align}
The bulk metric function $h(\eta)$ corresponds to the neural network weights $\theta$. 
To make the network depth finite, we introduce the UV and IR cutoffs for the metric as $\eta_{\rm ini}=1$, and $\eta_{\rm fin}=0.1$ in units of the AdS radius $L$. 

There are two big advantages of using Neural ODE. First, the metric function is smooth and we do not need to add penalty terms for smoothness. Therefore, we can largely reduce the number of hyper-parameters needed in the network. Second, our Neural ODE uses an adaptive ode solver, called ``dopri5.'' This gives us much more accuracy in the integration, and it turns out that the equation of motion in the curved geometry is sensitive to the discretization in some region of $\eta$. This adaptive method provides accuracy and efficiency simultaneously.

%%%%%%%%%%%%%%%%%%%%%%%%%%%%%%%%%%%%%%%%%%%%%%%%%%%%%%%%%%%%%%%%%%%%%%%%%%%%%%%%%%%%%%%%%
\subsubsection{Bulk metric parameterization}

To make the integration variable monotonically increase from the AdS boundary to the black hole horizon, we made a change of variable $\widetilde{\eta}=1-\eta$ for the metric function, and we model the metric function $h(\widetilde{\eta}$ using the following two ansatz:
\begin{align}
\text{ansatz 1: } &  h(\widetilde{\eta}) = \sum_{n=0}^{8}a_{n}\widetilde{\eta}^{n} \, ,
  \label{firstc}\\
\text{ansatz 2: }&  h(\widetilde{\eta}) = \sum_{n=0}^{8}b_{n}\widetilde{\eta}^{n}+\dfrac{1}{1-\widetilde{\eta}} \, .
  \label{secondc}
\end{align}
The first one is the Taylor series around the AdS boundary.
The second choice explicitly encodes the divergent behavior of the metric function near the black hole horizon at $\eta=0$. 
Any black hole horizon with a nonzero temperature has $f(\eta) \propto \eta^2$ with $g(\eta)$ being nonzero constant. Hence, Eq.~\eqref{hdef} leads to $h(\eta) \sim 1/\eta$ as the generic behavior of $h(\eta)$ near the horizon $\eta =0$. The second ansatz Eq.~\eqref{secondc} explicitly encodes this prior knowledge.

\begin{figure}
  \centering
  \includegraphics[width = 0.8\linewidth]{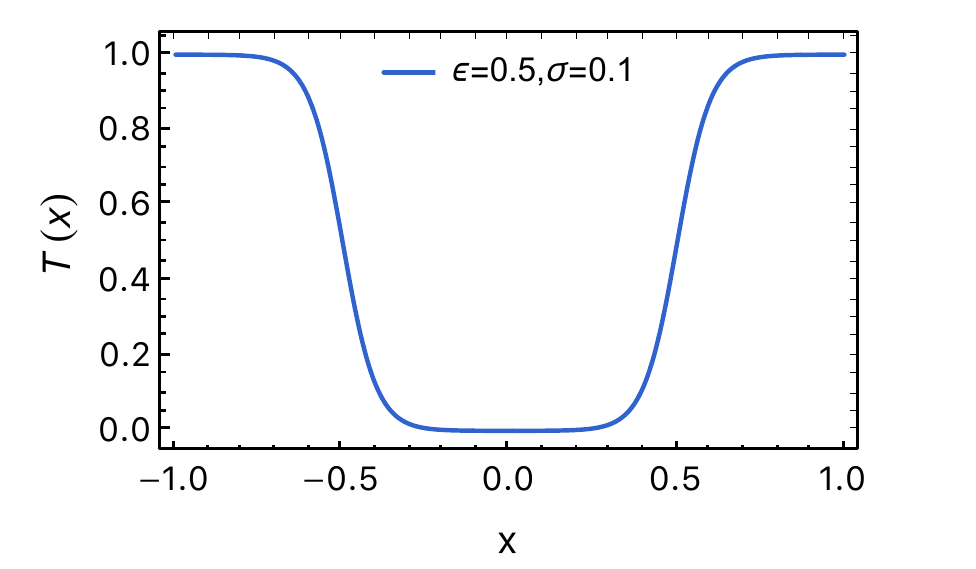}
  \caption{Differentiable nonlinear activation $T(x)$ given by Eq.~\eqref{Tx}.}
  \label{fig:Tx}
\end{figure}

%%%%%%%%%%%%%%%%%%%%%%%%%%%%%%%%%%%%%%%%%%%%%%%%%%%%%%%%%%%%%%%%%%%%%%%%%%%%%%%%%%%%%%%%%
\subsubsection{Lattice QCD data as input}
We use the lattice QCD data of RBC-Bielefeld collaboration \cite{phdthesisUnger} as our input data. 
The data is the chiral condensate $\mathcal{O}=\bar{q}q$, as a function of its source, the quark mass $m_{q}$. 
A plot is given in Fig.~\ref{fig:data} Left.
We take the $T=0.208$ [GeV] temperature data (the black line in Fig.~\ref{fig:data} Left), and the detail of the data is listed in Tab.~\ref{table:QCD_data}.\footnote{See \cite{Hashimoto:2018bnb} for the conversion method from the lattice QCD unit to the physical unit.}

We generate positive data and negative data in such a way that if the data's vertical distance to the experimental curve is less than 0.005, then it is labeled as positive (the label is 0). Otherwise, it is labeled as negative (the label is 1). We collected 10000 positive data and 10000 negative data used for training, as shown in Fig.~\ref{fig:data} Right.
%We consider a $\phi^{4}$ theory in the bulk, 
%\begin{equation}
%V[\phi] = \dfrac{\lambda}{4}\phi^{4}.
%\end{equation}
%Furthermore, the chiral condensate $\langle\bar{q}q \rangle$ has mass dimension 3 at the UV Gaussian fixed point. so we have to take $m^{2}=\dfrac{-3}{L^{2}}$\cite{Hashimoto:2018bnb}. 
Our goal is to obtain a holographic description of our QCD data using the Neural ODE method. The variation parameters are $\lambda$, $L$ and $h(\eta)$.

\begin{table}[t]
\caption{Chiral condensate as a function of quark mass \cite{phdthesisUnger}, at the temperature $T=0.208$ [GeV], converted to physical units \cite{Hashimoto:2018bnb}.}
\centering
\begin{tabular}{c c}
$m_{q}$[GeV] & $\langle\bar{\phi}\phi\rangle[({\rm GeV})^{3}]$ \\ [0.5ex] % inserts table %heading
\hline
\hline
0.00067 & 0.0063 \\
0.0013 & 0.012 \\
0.0027 & 0.021 \\
0.0054 & 0.038 \\
0.011 & 0.068\\
0.022 & 0.10 \\ [1ex]
\hline
\end{tabular}
\label{table:QCD_data}
\end{table}

\begin{figure*}[t]
    \centering
    \includegraphics[width = 0.9\linewidth]{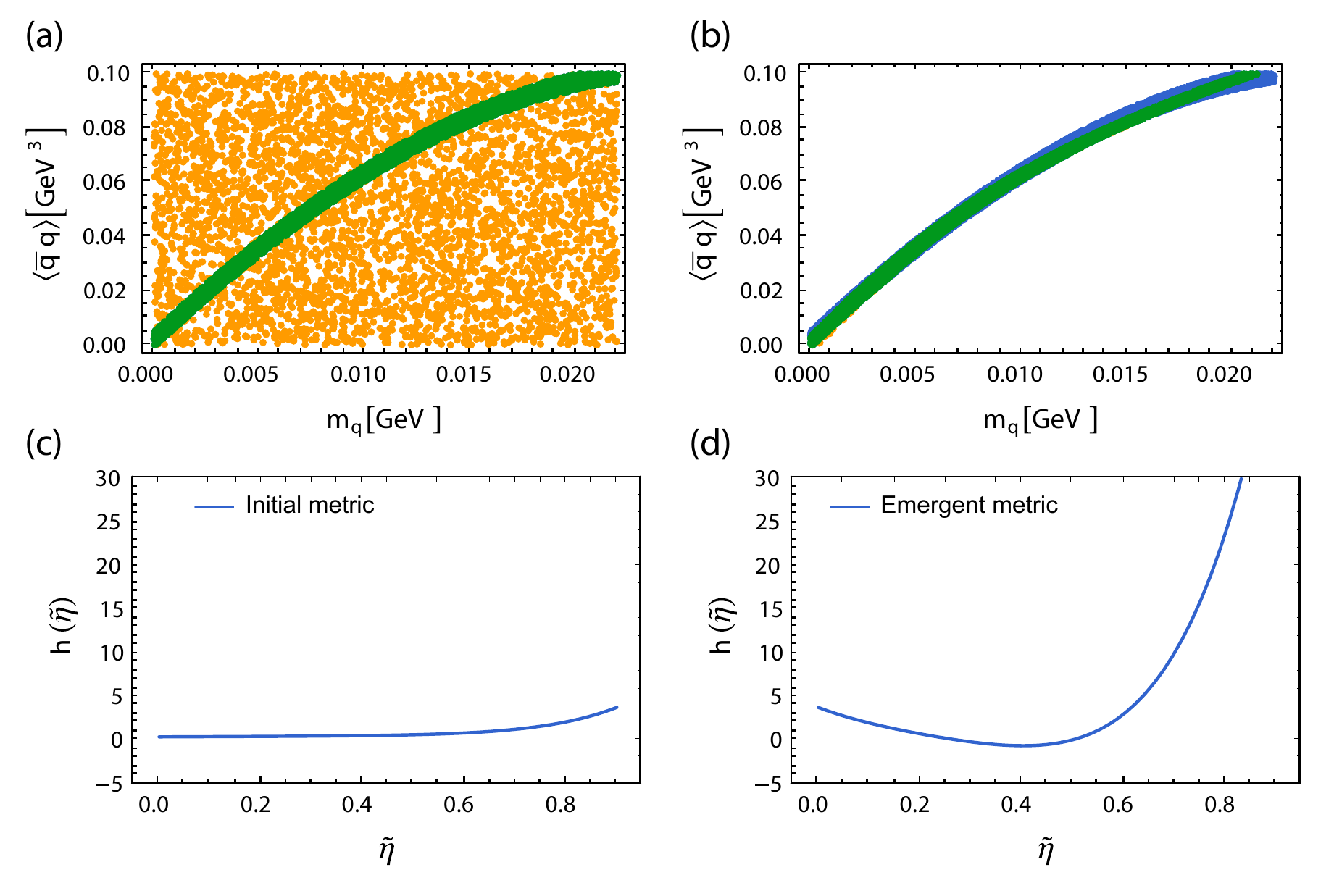}
    \caption{Subplot (a): The prediction of the machine before the training. Green dots are positive data. Orange dots are negative data but judged as positive by the machine as false positive. Subplot (b): the prediction of the machine after the training. Blue+green dots are positive data. Green dots are data judged as positive by the machine. Orange dots are the false positive data, which almost disappear after the training. Subplot (c): The metric function $h(\widetilde{\eta})$ before the training. The horizontal axis is $\widetilde{\eta}=1-\eta$. The black hole horizon is on the right side, $\widetilde{\eta}=1$, and the AdS boundary is on the left side, $\widetilde{\eta}=-\infty$. Subplot (d): the emergent metric $h(\widetilde{\eta})$ after the training. As we can see, the machine figures out the divergence behavior near the black hole horizon during the training.}
    \label{fig:res}
\end{figure*}

% \begin{figure*}[t]
% \centering
% \includegraphics[width = 0.8\linewidth]{init}
% \caption{Left: Green dots are positive data. Orange dots are negative data but judged as positive by the neural network.
% Right: The emergent metric $h(\eta)$ before the training. The horizontal axis is $\widetilde{\eta}\equiv 1-\eta$. The horizon is on the right side, $\widetilde{\eta}=1$, and the AdS boundary is on the left side, $\widetilde{\eta}=-\infty$. \label{fig:init}}
% \end{figure*}
% \begin{figure*}[t]
% \centering
% \includegraphics[width = 0.8\linewidth]{combined.pdf}
% \caption{
% Subplot (a): The prediction of the machine before the training. Green dots are positive data. Orange dots are negative data but judged as positive by the machine as false positive. Subplot (b): the prediction of the machine after the training. Blue+green dots are positive data. Green dots are data judged as positive by the machine. Orange data is the false positive date, which almost vanish after the training. Subplot (c): The emergent metric $h(\widetilde{\eta})$ before the training. The horizontal axis is $\widetilde{\eta}=1-\eta$. The black hole horizon is on the right side, $\widetilde{\eta}=1$, and the AdS boundary is on the left side, $\widetilde{\eta}=-\infty$. Subplot (d): the emergent metric $h(\widetilde{\eta})$ after the training. As we can see, the machine figures out the divergence behavior near black hole horizon during the training.
%  \label{fig:res}}
% \end{figure*}

%%%%%%%%%%%%%%%%%%%%%%%%%%%%%%%%%%%%%%%%%%%%%%%%%%%%%%%%%%%%%%%%%%%%%%%%%%%%%%%%%%%%%%%%%
\subsubsection{Loss function}

As for the loss function $\mathcal{L}$, we use \begin{equation}
\begin{split}
  \mathcal{L}=\frac{1}{N_{\text{data}}}
  \sum_{\rm data} &
  \left[
  \bigm|
  T(\pi(\eta_{\text{fin}});\epsilon,\sigma)-l\bigm|^2
  \right.
  \\&
  \left.+\beta\left(h(\eta_{\text{int}})-4\right)^{2}
  \right]
\end{split}
\label{loss}
\end{equation}
where the first term is the mean square error of the classifier loss function for the output data to approach the true result, Eq.~\eqref{blackhole_condition}.
The function $T(x;\epsilon,\sigma)$ is a specific differentiable nonlinear activation function that maps region $[-\epsilon,\epsilon]$ to 0, and otherwise to 1, in a fuzzy manner, 
\begin{equation}
T(x;\epsilon,\sigma)=1+0.5\left(\tanh\left(\dfrac{x-\epsilon}{\sigma}\right)-\tanh\left(\dfrac{x+\epsilon}{\sigma}\right)\right) \, .
\label{Tx}
\end{equation}
The parameter $\sigma$ controls the slope of the boundary as shown in Fig.~\ref{fig:Tx}. In the mean square error, $l$ is the label of the data ($l=0$ for positive data and $l=1$ for negative data). 
The second term in Eq.~\eqref{loss}, the $\beta$ penalty term, is to impose the condition that the emergent metric needs to be asymptotically AdS near the boundary $\eta=\eta_{\rm ini}$.
Due to nonlinear nature of the ODE function and sensitivity of Neural ODE, one may need to modify the hyperparameters $(\epsilon, \sigma)$ to ensure nonzero value of the gradient during the training.

%%%%%%%%%%%%%%%%%%%%%%%%%%%%%%%%%%%%%%%%%%%%%%%%%%%%%%%%%%%%%%%%%%%%%%%%%%%%%%%%%%%%%%%%%
\subsection{Emergent metric}

With the architecture described above, we perform the training. 
We first choose Eq.~\eqref{firstc} for the ansatz of the metric function $h(\eta)$. We randomly initialize the training parameters. 
The initial configuration of the metric function is given in the subplot (c) of Fig.\ref{fig:res}.
As shown in the subplot (a) of Fig.~\ref{fig:res}, the machine with the initial metric judges all the orange+green data as positive data. 

After training with 13000 epochs, the loss is reduced to 0.02. The result is shown in subplot (b) $\&$ (d) of Fig.~\ref{fig:res}. As we can see the predicted data agrees well with original positive data. We also observe that the emergent metric is a smooth function. The trained metric function reads:
\begin{eqnarray}
\begin{aligned}
h(\eta) = & \, 8.2352\widetilde{\eta}^{8} +8.0109\widetilde{\eta}^{7}+ 7.6072\widetilde{\eta}^{6} \\
& +6.9469\widetilde{\eta}^{5} + 150.89\widetilde{\eta}^{4} -130.81\widetilde{\eta}^{3} \\
& + 55.539\widetilde{\eta}^{2}-22.223\widetilde{\eta}^{1}+ 3.7720 \, .
\end{aligned}
\end{eqnarray}

The machine also finds the optimal values of the coupling constant and the AdS radius,
\begin{align}
& \lambda = 0.0004 \, , \\
& L = 5.1640 [{\rm GeV}^{-1}] \, .
\end{align}
As we can see in subplot (d) of Fig.~\ref{fig:res}, the metric function $h(\eta)$ which the Neural ODE found has tendency to grow significantly near $\eta \sim 0$. This is indeed the black hole horizon behavior. 
It is quite intriguing that the machine
automatically captures the divergence behavior of the metric function $h(\eta)$ near the black hole horizon. 

\begin{figure}
  \centering
  \includegraphics[width=0.8\linewidth]{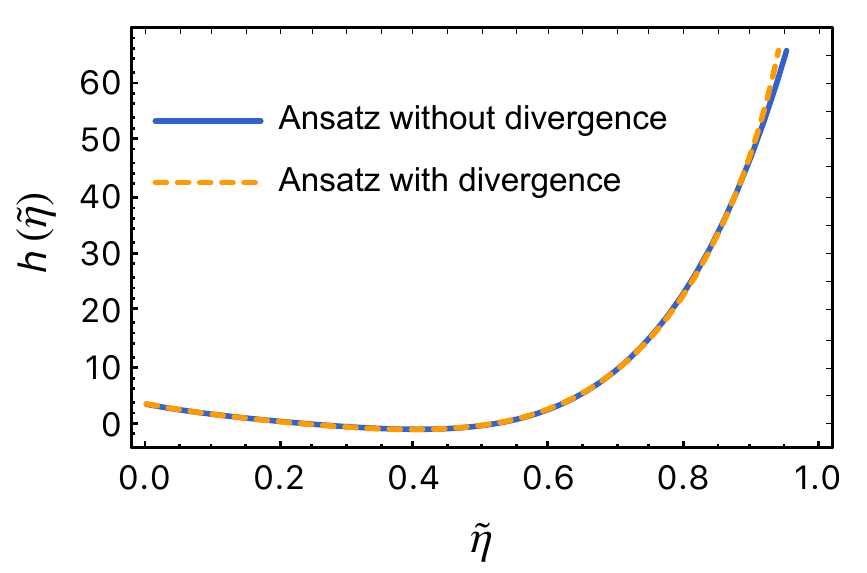}
  \caption{Two emergent metrics for $T=0.208$ [GeV] data with different metric ansatz. The solid (dashed) line is the trained result with Eq.~\ref{firstc} (Eq.~\ref{secondc}). The two lines are found to overlap with each other, thus the divergence behavior of metric function near the black hole horizon is emergent during the training.}
  \label{fig:emergent_metric}
\end{figure}

As a check, we also perform the training with the second ansatz for the metric function $h(\eta)$, {\it i.e.}~Eq.~\eqref{secondc}, which encodes the prior knowledge about the black hole horizon.
As shown in Fig.~\ref{fig:emergent_metric}, the result looks almost the same as that of the first ansatz that does not use the prior knowledge. Therefore, the regularization to implement the black hole horizon in $h(\eta)$ is not necessary. This result indicates that Neural ODE can automatically discover the black hole geometry in the holographic bulk and recover the near-horizon metric behavior without prior knowledge. For convenience, we use the training results of the second ansatz to calculate a physical observable (Wilson loop) in the next section.

%%%%%%%%%%%%%%%%%%%%%%%%%%%%%%%%%%%%%%%%%%%%%%%%%%%%%%%%%%%%%%%%%%%%%%%%%%%%%%%%%%%%%%%%%
\subsection{Multi-temperature result}

We also applied the above method to the multi-temperature QCD data given in Tab.~\ref{table:QCD_data_variousT}. 
During the training, we require different neural networks to share the same value of AdS radius $L$, and the training results are summarized in Tab.~\ref{table:multi-T}. 
The model discovers the optimal emergent metric as well as the coupling constant $\lambda$ at
each temperature. 

\begin{table*}[ht]
\caption{Chiral condensate as a function of quark mass \cite{phdthesisUnger}, at the values of temperature $T=0.188$, $0.192$, $0.196$, $0.200$, $0.204$, $0.208$ [GeV], converted to physical units \cite{Hashimoto:2018bnb}. The quark mass $m_q$ is in [GeV], and the chiral condensate $\langle\bar{\psi}\psi\rangle$ is in $[({\rm GeV})^{3}]$.}
\centering
\begin{tabular}{|c| c||c|c||c|c||c|c||c|c||c|c|}
\hline 
  \multicolumn{2}{|c||}{$T=0.188$} &
  \multicolumn{2}{c||}{$T=0.192$} &
  \multicolumn{2}{c||}{$T=0.196$} &
  \multicolumn{2}{c||}{$T=0.200$} &
  \multicolumn{2}{c||}{$T=0.204$} &
  \multicolumn{2}{c|}{$T=0.208$} 
  \\[0.5ex] \hline
$m_{q}$ & $\langle\bar{\psi}\psi\rangle$ &
$m_{q}$ & $\langle\bar{\psi}\psi\rangle$ &
$m_{q}$ & $\langle\bar{\psi}\psi\rangle$ &
$m_{q}$ & $\langle\bar{\psi}\psi\rangle$ &
$m_{q}$ & $\langle\bar{\psi}\phi\rangle$ &
$m_{q}$ & $\langle\bar{\psi}\psi\rangle$ 
\\ [0.5ex] 
\hline
\hline
0.00061	&0.056	&0.00062	&0.049	&0.00064	&0.034	&0.00065	&0.019	&0.00066	&0.011	&0.00068	&0.0064 \\
0.0012	&0.058	&0.0012	&0.053	&0.0013	&0.042	&0.0013	&0.027	&0.0013	&0.018	&0.0014	&0.012 \\
0.0024	&0.064	&0.0025	&0.059	&0.0025	&0.052	&0.0026	&0.040	&0.0026	&0.029	&0.0027	&0.022 \\
0.0049	&0.07	&0.005	&0.068	&0.0051	&0.065	&0.0052	&0.058	&0.0053	&0.048	&0.0054	&0.038 \\
0.0098	&0.08	&0.010	&0.081	&0.010	&0.081	&0.010	&0.079	&0.011	&0.075	&0.011	&0.068 \\
0.020	&0.095	&0.020	&0.098	&0.020	&0.10	&0.021	&0.10	&0.021	&0.10	&0.022	&0.10 \\
[1ex]
\hline
\end{tabular}
\label{table:QCD_data_variousT}
\end{table*}

\begin{table*}[ht]
\caption{Left (Right) : Multi-temperature result for the metric without (with) the divergence ansatz.}
\centering
\begin{tabular}{|c||c|c|c|c|c|c|}
\hline
\centering
\small
\setlength\tabcolsep{3pt}

 $T$ & 0.188 & 0.192 & 0.196 & 0.200 & 0.204 & 0.208\\ % inserts table
%heading
\hline

$L$ &5.164 &5.164&5.164&5.164&5.164&5.164 \\ 
\hline
$\lambda$ & 0.0014 & 0.0011 & 0.0009 & 0.0007 & 0.0005 & 0.0003\\ \hline
$a_0$ & 3.7671 & 3.7678 & 3.7688 &3.7698 & 3.7709 & 3.7720\\ \hline

$a_1$ & -22.229 &-22.228& -22.227&-22.226 & -22.225 & -22.223\\
\hline
$a_2$ &55.533 & 55.534 & 55.535 & 55.536 & 55.537 & 55.539\\
\hline
$a_3$ &-130.82 & -130.82 & -130.81 & -130.81 &-130.81 & -130.81 \\
\hline
$a_4$ & 150.88 &150.88 & 150.88 &150.88& 150.88 & 150.89\\
\hline
$a_5$ & 6.939 & 6.9424 & 6.9434 & 6.9443 & 6.9457 & 6.9469\\
\hline
$a_6$ &7.5981 & 7.6026 & 7.6036 & 7.6044 & 7.6061 & 7.6072\\
\hline
$a_7$ & 8.0004 & 8.0062 & 8.0071 & 8.0079 & 8.0098 & 8.0109\\
\hline
$a_8$ & 8.2230 &8.2304 & 8.2313 & 8.2320 & 8.2341 & 8.2352\\
\hline

\end{tabular}
%\label{table:multi-T-without-divergence}
%\end{table*}
%
%
%\begin{table*}[ht]
%\caption{Multi-temperature result(metric with divergence)}
%\centering
\hspace*{5pt}
\begin{tabular}{|c||c|c|c|c|c|c|}
\hline
\centering
\small
\setlength\tabcolsep{3pt}
 $T$ & 0.188 & 0.192 & 0.196 & 0.200 & 0.204 & 0.208\\ 
\hline

$L$ &5.164 &5.164&5.164&5.164&5.164&5.164 \\ 
\hline
$\lambda$ & 0.0014 & 0.0011 & 0.0009 & 0.0007 & 0.0005 & 0.0003\\ \hline
$b_0$ & 2.8430 & 2.8438 & 2.8447 & 2.8456 & 2.8467 & 2.8474\\ \hline

$b_1$ & -24.140 & -24.139 & -24.138 &-24.137 & -24.136 & -24.135\\
\hline
$b_2$ &55.627 & 55.628 & 55.629 & 55.630 & 55.631 & 55.632\\
\hline
$b_3$ &-130.22 & -130.22& -130.22 & -130.22 &-130.22 & -130.22 \\
\hline
$b_4$ & 150.79 & 150.79 & 150.79 & 150.79 & 150.79 & 150.80\\
\hline
$b_5$ & 5.5746 & 5.5774 & 5.5790 & 5.5802 & 5.5813 & 5.5820 \\
\hline
$b_6$ &4.6816 & 4.6849 & 4.6867 & 4.6880 & 4.6891 & 4.6898 \\
\hline
$b_7$ & 3.5672 & 3.5710 & 3.5730 & 3.5744 & 3.5756 & 3.5763 \\
\hline
$b_8$ & 2.5329 & 2.5371 & 2.5394 & 2.5409 & 2.5421 & 2.5428\\
\hline

\end{tabular}
\label{table:multi-T}
\end{table*}

We have two observations of the trained results shown in Tab.~\ref{table:multi-T}.
First, the obtained metric $h(\eta)$ and the AdS radius $L$ do not depend on the temperature $T$.
Second, the only dependence on the temperature is encoded solely in the coupling constant $\lambda$
of the scalar field theory.

The former sounds counter-intuitive, since normally the metric itself should be highly dependent on the temperature,
and the change in the metric will modify the gravitational fluctuation, which corresponds to the gluon physics. 
It is easy to resolve this issue. The obtained function is $h(\eta)$ and not the full metric components $f(\eta)$ 
and $g(\eta)$. Even for the case of the AdS Schwarzschild geometry in which
the metric is temperature-dependent,
we find $h(\eta)=\frac{4}{L} \coth \frac{4\eta}{L}$ which is temperature independent.
In the next section, to compute physical quantities from the emergent $h(\eta)$, we assume some functional form of $g(\eta)$ and discuss the temperature dependence of the metric components.

What the machine found is that the reproduction of the input data mainly relies on the temperature dependence of the coupling constant $\lambda$ in the holographic bulk theory.
For lower temperature, we find a strong nonlinear interaction, {\it i.e.} larger $\lambda$.
The value of $\lambda$ is directly related to the self-coupling of sigma meson. Although we cannot compare our trained results with experiments since the self-coupling has never been precisely measured due to the broad width of the sigma meson, our result provides a unique view of the QCD phase transition, in particular about the mysterious relation between the chiral transition and the deconfinement transition.

%%%%%%%%%%%%%%%%%%%%%%%%%%%%%%%%%%%%%%%%%%%%%%%%%%%%%%%%%%%%%%%%%%%%%%%%%%%%%%%%%%%%%%%%%
\section{Physical interpretation of the emergent spacetime}
\label{sec:4}

%%%%%%%%%%%%%%%%%%%%%%%%%%%%%%%%%%%%%%%%%%%%%%%%%%%%%%%%%%%%%%%%%%%%%%%%%%%%%%%%%%%%%%%%%
\subsection{Reconstruction of the metric}

Since in our case the machine learns only $h(\eta)$, to compute physical quantities such as Wilson loop, we need
to assume the form of $g(\eta)$ to get $f(\eta)$. Here we assume 
the functional form of the AdS Schwarzschild configuration, 
\begin{align}
g(\eta) = A \left(\cosh\frac{2\eta}{La}\right)^a \, ,
\label{gansatz}
\end{align}
where $A$ and $a$ are temperature-dependent constant. In particular the constant $a$ encodes the dimensionality of the AdS${}_{d+1}$-Schwarzschild as $a = d/4$, and here we just set it as a free parameter.
The ansatz Eq.~\eqref{gansatz} 
also satisfies the criterion that $g$ is a monotonic function of $\eta$, which is normally required for 
spacetimes without a bottle neck.
The Hawking temperature $T$ constrains the function $f(\eta)$ as
\begin{align}
f(\eta) \sim (2\pi T)^2 \eta^2
\end{align}
so, for our calculation we define a new function $F(\eta)$ as
\begin{align}
f(\eta) = (2\pi TL)^2 \left(\tanh \eta/L\right)^2 F(\eta)
\label{Fdef}
 \end{align}
which satisfies the boundary condition
\begin{align}
\lim_{\eta \to 0} F(\eta) = 1 \, .
\label{F0}
\end{align}
Substituting Eqs.~\eqref{gansatz} and \eqref{Fdef} to Eq.~\eqref{hdef}, and perform the integration over $\eta$
with the integration constant fixed by Eq.~\eqref{F0}, we obtain
\begin{align}
F(\eta) = \exp \int_0^\eta \left(
2h(\eta) - \frac{4}{L\sinh (2\eta/L)}-\frac{6}{L} \tanh\frac{2\eta}{La}
\right) d\eta \, .
\label{Fres} 
\end{align}
The overall factor $A$ in $g(\eta)$ in Eq.~\eqref{gansatz} can be fixed by the following
asymptotically AdS${}_5$ constraint at $\eta \gg L$, 
\begin{align}
f(\eta) \simeq g(\eta) \simeq e^{2\eta/L + \mbox{const.}}, 
\label{AsAdS}
\end{align}
which implies $h(\eta)\simeq 4/L$ according to Eq.~\eqref{hdef}.
To determine this constant which we require temperature independent, 
we expand Eq.~\eqref{Fres} around $\eta \gg L$ as
\begin{align}
&\int_0^\eta \left(
 2h(\eta) - \frac{4}{L\sinh (2\eta/L)}-\frac{6}{L} \tanh\frac{2\eta}{La}
\right) d\eta 
\nonumber \\
& = \frac{2\eta}{L} + c(a) + {\cal O}(1/\eta) \, .
\end{align}
Using this constant $c(a)$, 
the constraint Eq.~\eqref{AsAdS} determines the normalization of $g(\eta)$ as
\begin{align}
g(\eta) = (2\pi T L)^2 e^{c(a)} \left(2\cosh\frac{2\eta}{La}\right)^a \, .
\label{gdet}
\end{align}
Now, since we require that the constant in Eq.~\eqref{AsAdS} is temperature independent,
we have a condition
\begin{align}
\frac{\partial}{\partial T} \left[ T^2 e^{c(a(T))} \right] = 0 \, .
\end{align}
Up to an integration constant, we can numerically solve this equation.
Assuming that at $T=0.208$ [GeV] we have $a=1$, we find numerically $c(a=1) = 11.1952$. 
Then the equation above leads to $c(a(T=0.188 \, {\rm [GeV]})) = 11.3984$ and 
$a(T=0.208 \, {\rm [GeV]})=1.098$. We are going to use $g(\eta)$ given by Eq.~\eqref{gdet} and 
$f(\eta)$ given by Eq.~\eqref{Fdef} with Eq.~\eqref{Fres} for the calculation of physical quantities
below.

\begin{figure*}
\centering
\includegraphics[width = 0.45 \linewidth]{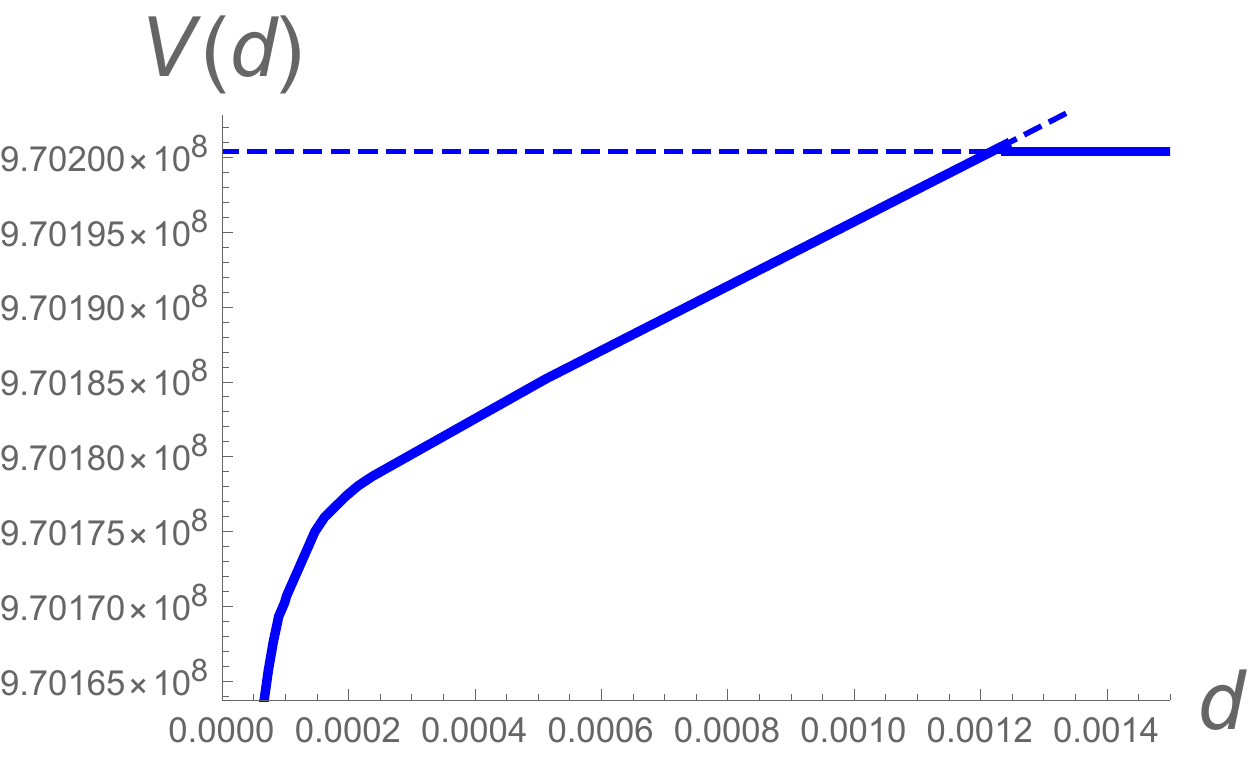}
\includegraphics[width = 0.45 \linewidth]{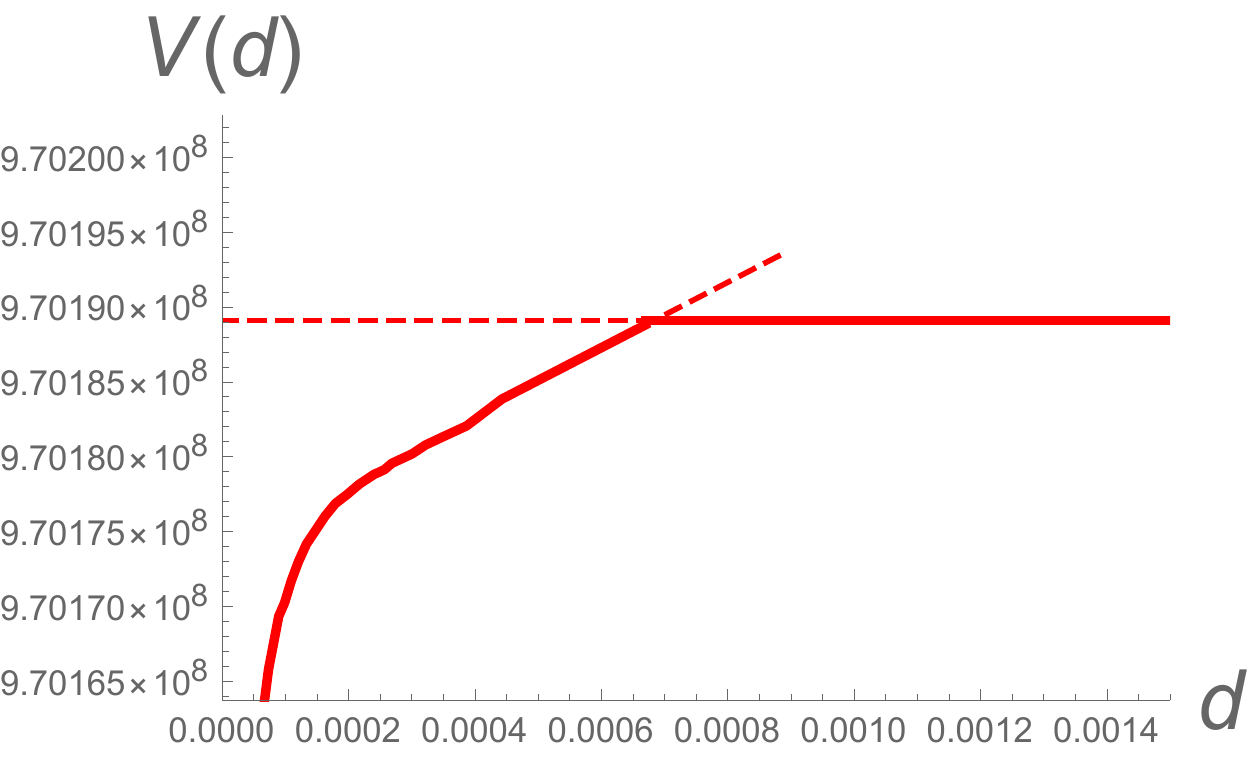}
\caption{The quark antiquark potential $V(d)$. Left: calculated in the emergent spacetime with the metric functions
trained with the
data at $T=0.188$ [GeV]. Right: that at $T=0.208$ [GeV].
In these figures, the zeros of the vertical axis should be ignored, as they are dependent on the cut-off of the asymptotic AdS boundary.
\label{fig:Wilson}}
\end{figure*}

%%%%%%%%%%%%%%%%%%%%%%%%%%%%%%%%%%%%%%%%%%%%%%%%%%%%%%%%%%%%%%%%%%%%%%%%%%%%%%%%%%%%%%%%%
\subsection{Wilson loop}

Following the standard method \cite{Maldacena:1998im,Rey:1998ik,Rey:1998bq}
for calculating the expectation value of 
the Wilson loop holographically, 
we evaluate the Wilson loop
for a quark and an antiquark separated by the distance $d$, using our emergent spacetime.
The logarithm of the Wilson loop $\langle W \rangle$, which is proportional to the quark potential $V$,  
is the area of the Euclidean worldsheet of a string hanging down from the AdS boundary.
The string reaches $\eta=\eta_0$ at the deepest, and both the quark potential $V(d)$ and the quark distance
$d$ are functions of $\eta_0$, as
\begin{align}
d = 2\int_{\eta_0}^\infty
\frac{1}{\sqrt{g(\eta)}}\sqrt{\frac{f(\eta_0)g(\eta_0)}{f(\eta)g(\eta)-f(\eta_0)g(\eta_0)}}d\eta \, ,
\label{d}
\\
2\pi \alpha' V = 2\int_{\eta_0}^\infty \!\!\!\!
\sqrt{f(\eta)}
\sqrt{\frac{f(\eta_0)g(\eta_0)}{f(\eta)g(\eta)-f(\eta_0)g(\eta_0)}}d\eta \, .
\label{Vd}
\end{align}
Here $1/(2\pi \alpha')$ is the string tension which is undetermined in this work.
Eliminating $\eta_0$ from these expressions implicitly defines $V(d)$.
Note that the integration in $V(d)$ diverges at $\eta = \infty$, and we need to introduce 
a cut-off for the asymptotic AdS boundary for the calculation.

The quark potential $V(d)$ has another saddle, which is just two straight strings connecting the
black hole horizon and the asymptotic boundary,
\begin{align}
2\pi \alpha' V_{\rm Debye}
=2\int^{\eta_0}_0 \!\!\!\!
\sqrt{f(\eta)}d\eta\, .
\label{V}
\end{align}
We need to adopt $V(d)$ in Eq.~\eqref{Vd} or $V_{\rm Debye}$, whichever is smaller.

\begin{figure*}
\centering
\includegraphics[width = 0.45 \linewidth]{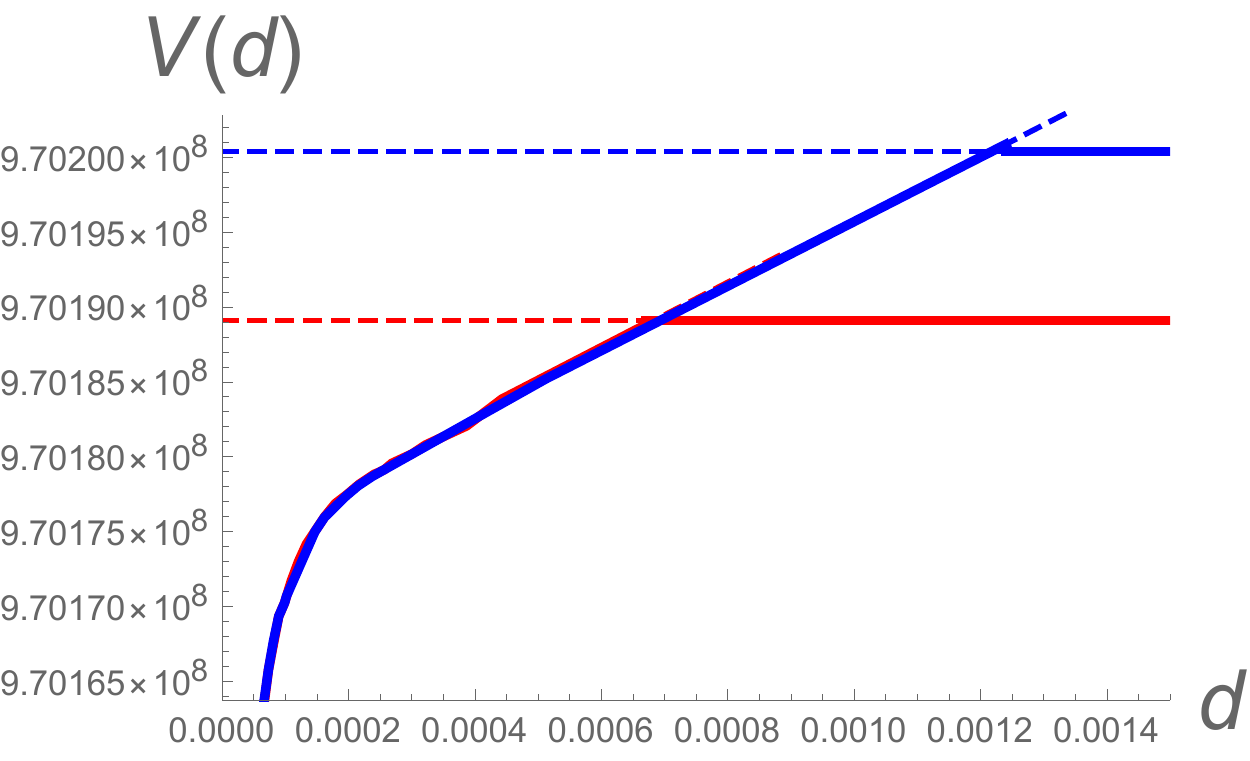}
\includegraphics[width = 0.45\linewidth]{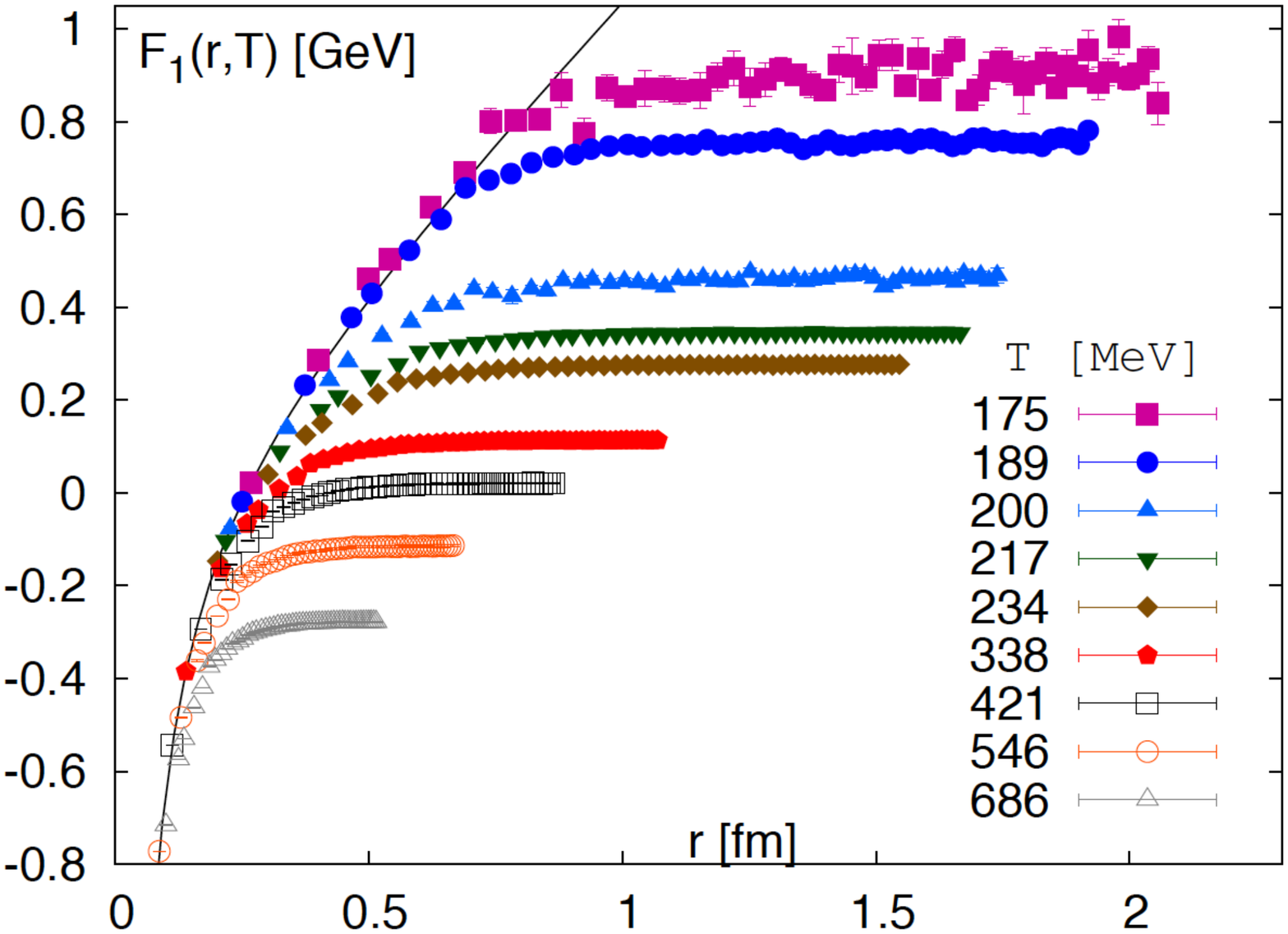}
\caption{
Left: calculated quark-antiquark potential for $T=0.188$ [GeV] (blue line) and for $T=0.208$ [GeV] (red line).
Two lines overlap with each other, except for the flat parts.
Right: lattice QCD result of the quark-antiquark potential at different values of temperature, taken from Ref.~\cite{Petreczky:2010xg}.
\label{fig:lattice}}
\end{figure*}

Using the metric obtained in the previous subsection, we calculate the quark potential for each temperature.
In Fig.~\ref{fig:Wilson}, we present the quark potential for $T=0.188$ [GeV] data and $T=0.208$ [GeV] data.
They exhibit three phases: at short $d$, the potential is Coulombic, while at large $d$, the potential is
flat and Debye-screened, and in the middle range of $d$, the potential is linear, signifying the quark confinement.
The set of these features is well-known in lattice QCD simulations (see Fig.~\ref{fig:lattice}), and, interestingly, 
our holographic results reproduce these features.\footnote{In Ref.~\cite{Andreev:2006nw}, phenomenological ansatz for
the bulk spacetime which is similar to ours and that of Ref.~\cite{Hashimoto:2018bnb} was made.
Surprisingly, the machine learns a metric that was proposed independently by humans. K.H.~would like
to thank Oleg Andreev for bringing the paper to his attention.}

This reproduction was reported in Ref.~\cite{Hashimoto:2018bnb}, and here we further investigate the 
temperature dependence.
As we see in Fig.~\ref{fig:Wilson}, the two plots are identical with each other except for the height of the
Debye screening parts. The higher temperature corresponds to the lower height of the flat potential,
which is qualitatively consistent with the lattice QCD result, as shown in Fig.~\ref{fig:lattice}.

%%%%%%%%%%%%%%%%%%%%%%%%%%%%%%%%%%%%%%%%%%%%%%%%%%%%%%%%%%%%%%%%%%%%%%%%%%%%%%%%%%%%%%%%%
\section{Summary and discussion}
\label{sec:5}

In this paper, we applied the Neural ODE to the AdS/DL correspondence, where the emergent spacetime in the gravity side of the AdS/CFT correspondence is regarded as a deep neural network.
Since the classical spacetime is continuous and smooth, the weights of the network need to be interpreted as a smooth function of the depth, thus the Neural ODE provides a very natural scheme for training the bulk geometry. We followed the setup of Ref.~\cite{Hashimoto:2018bnb} of using the lattice QCD data of 
QCD chiral condensate to train the neural network. We demonstrated that the Neural ODE indeed worked well to discover a bulk geometry which is holographically consistent with
the lattice QCD data. Even without including the black hole boundary condition for the ansatz function of the Neural ODE, the machine found automatically the black hole horizon behavior. This proves the ability of the Neural ODE to automate the proposal of the holographic bulk theory from the holographic boundary data in the AdS/CFT setup.

We performed the training with the training data of lattice QCD at various temperatures and found that the optimal volume factor of the emergent geometries shares the same radial dependence except for the overall normalization. The temperature dependence in the behavior of the QCD chiral condensate simply comes from the bulk scalar coupling constant, which corresponds to the meson couplings.
The Wilson loops holographically calculated with the machine-trained emergent geometries appeared to have a correct temperature dependence, as in Fig.~\ref{fig:lattice}.

For a more quantitative evaluation of the emergent spacetime, here we argue that the slope of the linear part of the plots of the quark-antiquark potential,
given in Fig.~\ref{fig:Wilson},
corresponds to the QCD string tension $\sigma$. Since in our formulation, the overall
normalization $2\pi \alpha'$ is not given, 
we only look at the ratio of the slope at $T=0.188$ [GeV]
and the slope at $T=0.208$ [GeV]. A numerical fitting of Fig.~\ref{fig:Wilson}
gives $\sigma_{T=0.208{\rm GeV}}/\sigma_{T=0.188{\rm GeV}} \simeq 1.0$. 
In lattice QCD simulation, this number is expected to be smaller than $1$,
because the deconfinement transition (which is not the first-order phase transition) 
occurs when the QCD string tension goes to zero.
So our value $1.0$ still keeps the tendency of the large $N$ gauge theories
where the deconfinement transition is expected to be the first order.

In addition, we notice that the string breaking distance, the value of $d$ at 
the kink in Fig.~\ref{fig:Wilson}, is around $d \sim 10^{-3}$ in the unit of
$L \sim 5$ [GeV${}^{-1}$], which is too small
compared to the expected QCD value $d\sim {\cal O}(1)$[fm]. 
This quantitative discrepancy would be largely due to our assumed functional form 
of the metric component $g(\eta)$ in Eq.~\eqref{gansatz}.
In this paper we have seen the qualitative feature of the temperature dependence
of the Wilson loops to be consistent with lattice QCD results\footnote{In fact, we required that the constant in Eq.\ref{AsAdS} is independent of the temperature, and if we loosen this condition, the resultant Wilson loops do not match the lattice QCD results.}, and further quantitative
match will need some different observable data to train $f(\eta)$ and $g(\eta)$ independently.

The Neural ODE is quite effective for physical applications of the machine learning method in which neural network weights have physical meanings. Any physical observable, if looked minutely enough, should be a continuous function of space and time. To identify weights of standard deep neural networks with physical quantities, regularizations to make them a smooth function on the discrete network are necessary, which are rather artificial and often still can not remove discretization artifacts fully. In Neural ODEs, the weights are continuous functions in the first place, which hence reduces unnecessary ingenuity of the regularizations. 
One of the main improvements from Ref.~\cite{Hashimoto:2018bnb}, although the physical setup is the same, is that we could remove the artificial regularizations used in \cite{Hashimoto:2018bnb}, and largely improve the prediction accuracy of the emergent bulk metric at the same time. 

Since we obtained the emergent volume factor for each temperature, it is possible to ask what kind of bulk action can allow such a metric as a solution of its equation of motion. There is a lot of work that elaborated possible bulk systems dual to QCD, and the major example would be the Einstein-dilaton system \cite{Gursoy:2007cb,Gursoy:2007er}. We want to visit this question in future publications.

%%%%%%%%%%%%%%%%%%%%%%%%%%%%%%%%%%%%%%%%%%%%%%%%%%%%%%%%%%%%%%%%%%%%%%%%%%%%%%%%%%%%%%%%%
\acknowledgments

We would like to thank T.~Akutagawa and T.~Sumimoto for valuable discussions. We thank Microsoft Research for the kind hospitality during the workshop ``Physics $\cap$ ML.'' H.-Y.~H.~would like to thank Lei Wang for the discussion on Neural ODE. 
K.~H.~was supported in part by JSPS KAKENHI Grant Number JP17H06462. H.-Y.~H.~and Y.-Z.~Y.~were supported by a startup fund from UCSD.

%%%%%%%%%%%%%%%%%%%%%%%%%%%%%%%%%%%%%%%%%%%%%%%%%%%%%%%%%%%%%%%%%%%%%%%%%%%%%%%%%%%%%%%%%
\appendix

\section{Neural ODE}
\label{sec:A}
In this appendix, we briefly introduce Neural ODE \cite{chen2018neural}, and how to backpropagate the errors to train parameters. We assume the dynamics of a set of variables $\vec{x}(t)=\{x_{i}(t)\}$ can be described by the ODE specified by a velocity function $\vec{v}=\{v_{i}(\vec{x}(t),t;\theta)\}$, where $\theta$ are training parameters. We call the following equation the forward ODE,
\begin{equation}
    \dfrac{dx_{i}(t)}{dt}=v_{i}(\vec{x}(t),t;\theta) \, .
\end{equation}
Given the initial condition $x_i(0)$, the ODE can be integrated from $t=0$ to $t=1$. The loss function $\mathcal{L}$ is a function of the final state,
\begin{equation}
    \mathcal{L}=\mathcal{L}(\vec{x}(1)).
\end{equation}
To calculate the gradient with respect to the parameter $\theta$, we first need to calculate the gradient with respect to $\vec{x}(t)$ at each time t. Define the adjoint variable $\vec{a}(t)=\{a_{i}(t)\}$
\begin{equation}
    a_{i}(t)=\dfrac{\partial \mathcal{L}}{\partial x_{i}(t)} \, .
\end{equation}
To derive the dynamics of adjoint variables, we consider the dependence chain $\vec{x}(t)\rightarrow \vec{x}(t+dt)\rightarrow \cdots \rightarrow \mathcal{L}$,
\begin{equation}
    \dfrac{\partial \mathcal{L}}{\partial x_i(t)}=\dfrac{\partial \mathcal{L}}{\partial x_{j}(t+dt)}\dfrac{\partial x_j(t+dt)}{\partial x_{i}(t)} \, ,
\end{equation}
where Einstein summation is assumed. Then we find
\begin{equation}
\begin{split}
    a_{i}(t)&=a_j(t+dt)\dfrac{\partial [x_{j}(t)+v_{j}(\vec{x}(t),t;\theta)dt]}{\partial x_i(t)}\\
    &=(\delta_{ij}+\partial_{x_i(t)}v_{j}(\vec{x}(t),t;\theta)dt)a_j(t+dt) \, .
\end{split}
\end{equation}
Therefore, the adjoint variable follows backward ODE equation,
\begin{equation}
    \begin{split}
        \dfrac{da_i(t)}{dt}=-a_{j}(t)\partial_{x_i(t)}v_{j}(\vec{x}(t),t;\theta)\, ,\label{eq:backprop}
    \end{split}
\end{equation}
\begin{equation}
    a_i(0)=\int^{0}_{1}a_j(t)\partial_{x_i(t)}v_{j}(\vec{x},t;\theta)dt \, .
\end{equation}
To calculate the gradient with respect to the parameter $\theta$, we can collect the gradient for each time step backward,
\begin{equation}
\begin{split}
    \dfrac{\partial \mathcal{L}}{\partial \theta}&=\int^{0}_{1}\dfrac{\partial \mathcal{L}}{\partial x_i(t)}\dfrac{\partial(x_i(t)-x_i(t-dt))}{\partial \theta}\\
    &=\int^{0}_{1}\dfrac{\partial \mathcal{L}}{\partial x_i(t)}\dfrac{\partial v_i(\vec{x},t;\theta)}{\partial \theta}dt \, .
\end{split}
\end{equation}

%\section{Multi-temperature training result}
%\label{sec:B}

%Here, we listed the training results with the following metric ansatz in table.\ref{table:multi-T-with-divergence}.
%\begin{equation}
%    h(\eta)=\sum_{n=0}^{8}a_{n}\widetilde{\eta}^{n}+\dfrac{1}{1-\widetilde{\eta}}.
%\end{equation}{}

%%%%%%%%%%%%%%%%%%%%%%%%%%%%%%%%%%%%%%%%%%%%%%%%%%%%%%%%%%%%%%%%%%%%%%%%%%%%%%%%%%%%%%%%%
\bibliography{holographic_QCD}

\end{document}